\documentclass[12pt,letter]{article}

\usepackage{scicite,Abbrev}

\usepackage{times}

\usepackage{graphicx}

\newenvironment{sciabstract}{%
\begin{quote} \bf}
{\end{quote}}

\newcommand{\sg}{\delta_{\rm SG}}
\newcommand{\si}{\delta_{\rm SI}}
\newcommand{\gi}{\delta_{\rm GI}}
\newcommand{\di}{\delta_{\rm DI}}
\newcommand{\sstar}{\sigma^\star/m}
\newcommand{\csg}{{\rm cm}^2/{\rm g}}
\newcommand{\sdm}{\sigma_{\rm DM}/m}
\def\simlt{\lower.5ex\hbox{$\; \buildrel < \over \sim \;$}}
\def\simgt{\lower.5ex\hbox{$\; \buildrel > \over \sim \;$}}

\def\papertitle{The non-gravitational interactions of dark matter in colliding galaxy clusters}
\title{\papertitle} 

\author
{David Harvey$^{1,2\ast}$, Richard Massey$^{3}$, Thomas Kitching$^{4}$,\\
Andy Taylor$^2$, Eric Tittley$^2$
\\
\normalsize{$^{1}$Laboratoire d'astrophysique, EPFL, Observatoire de Sauverny, 1290 Versoix, Switzerland}\\
\normalsize{$^{2}$Royal Observatory, University of Edinburgh, Blackford Hill, Edinburgh EH9 3HJ, UK}\\
\normalsize{$^{3}$Institute for Computational Cosmology, Durham University, South Road, Durham DH1 3LE, UK}\\
\normalsize{$^{4}$Mullard Space Science Laboratory, University College London, %Holmbury St Mary, 
Dorking, Surrey RH5 6NT, UK}
\\
\normalsize{$^\ast$To whom correspondence should be addressed; E-mail: david.harvey@epfl.ch}
}

\date{}

\begin{document} 


\baselineskip24pt


\maketitle


\begin{sciabstract}
Collisions between galaxy clusters provide a test of the non-gravitational forces acting on dark matter. 
Dark matter's lack of deceleration in the `bullet cluster collision' constrained its self-interaction cross-section $\sdm<1.25\,\csg$ (68\% confidence limit) for long-ranged forces.
%if the scattering is anisotropic; 
%its lack of mass loss constrained $\sdm<0.7\csg$ (68\% CL) 
%for short-ranged forces. 
%if the scattering is isotropic.
%Such constraints are limited by uncertainty in the impact velocity, impact parameter and 3D orientation with respect to the line of sight of individual systems \cite{bulletcluster}. 
%However, all galaxy clusters grow through almost continual minor merger accretion, which can be exploited statistically \cite{bulleticity,Harvey14}. 
Using the Chandra and Hubble Space Telescopes we have now observed 72 collisions, including both `major' and `minor' mergers.
Combining these measurements statistically, we detect the existence of dark mass at $7.6\sigma$ significance.
The position of the dark mass has remained closely aligned within $5.8\pm8.2$\,kpc of associated stars: implying a 
%, and we constrain its
self-interaction cross-section $\sdm<0.47\,\csg$ (95\% CL) and disfavoring some proposed extensions to the standard model.
%Collisions between galaxy clusters provide a test of the forces acting on dark matter. 
%The apparent lack of deceleration of dark matter in the `bullet cluster' constrains its self-interaction cross-section $\sdm<1.25\,\csg$  (68\%CL) \cite{impactpars} for long-range, non-gravitational forces \cite{SIDMModel}. 
%Such constraints are limited by uncertainty in the impact velocity, impact parameter and 3D orientation with respect to the line of sight of individual systems \cite{bulletcluster}. 
%However, all galaxy clusters grow through almost continual minor merger accretion, which can be exploited statistically \cite{bulleticity,Harvey14}. 
%Using the Chandra and Hubble Space Telescopes, we observe 72 merging events, both large and small. 
%We detect the existence of dark mass at $7.6\sigma$ statistical significance, and constrain its angle-averaged momentum transfer cross-section $\sdm<0.47\,\csg$ (95\% CL): disfavouring some extensions to the standard model \cite{SIMP}.
\end{sciabstract}

% In setting up this template for *Science* papers, we've used both
% the \section* command and the \paragraph* command for topical
% divisions.  Which you use will of course depend on the type of paper
% you're writing.  Review Articles tend to have displayed headings, for
% which \section* is more appropriate; Research Articles, when they have
% formal topical divisions at all, tend to signal them with bold text
% that runs into the paragraph, for which \paragraph* is the right
% choice.  Either way, use the asterisk (*) modifier, as shown, to

%SCARY ENGLISH ("PARADIGMS") FOR A FIRST SENTENCE! REWRITTEN TO BE LESS OFFPUTTING
%Many studies have now provided evidence to support a paradigm in which most of the matter in the Universe lies beyond the standard model of particle physics. 
Many independent lines of evidence now suggest that most of the matter in the Universe is in a form outside the standard model of particle physics. % \cite{planck14}. 
A phenomenological model for cold dark matter \cite{EvolutionLSS} has proved hugely successful on cosmological scales,
%ADDED A LINE
where its gravitational influence dominates the formation and growth of cosmic structure. However,
% but 
there are several challenges on smaller scales: the model incorrectly predicts individual galaxy clusters to have more centrally concentrated density profiles \cite{corecusp}, larger amounts of substructure \cite{lostsatellite1,lostsatellite}, and the Milky Way to have more satellites able to produce stars \cite{toobigtofail} than are observed. These inconsistencies could be resolved through astrophysical processes \cite{baryonsolution}, or if dark matter particles are either warm \cite{GaussianStats} or self-interact with cross-section $0.1\le\sdm\le1\,\csg$ \cite{ObserveSIDM,SIDMSimA,SubhalosSIDMA}.
%ADDED CAVEAT HERE TO SAVE SANITY LATER (EVERYTHING ELSE UNDER THE SUN NOW NEEDS TO BE DEFINED IN THE PENULTIMATE PARAGRAPH!)
Following \cite{SIDMModel}, we define the momentum transfer per unit mass $\sdm$,
%as in equation 8 of ref.\ \cite{SIDMModel}, 
integrating over all scattering angles and assuming that individual dark matter particles are indistinguishable.

Self-interaction within a hidden dark sector is a generic consequence of some extensions to the standard model. 
%For example, mirror dark matter \cite{mirror}, glueball \cite{glueballs}, or SIMP \cite{SIMP} models with appropriate particle masses 
For example, models of mirror dark matter  \cite{mirror} and hidden sector dark matter \cite{mirror,glueballs,SIMP,composite,resonant }
all predict anisotropic scattering with %a velocity-independent cross-section 
$\sdm\approx1\,\rm{barn/GeV}=0.6\,\csg$, 
similar to nuclear cross-sections in the standard model. 
Note that couplings within the dark sector can be many orders of magnitude larger than those between dark matter and standard model particles, which is at most of order picobarns \cite{LUX}.

In terrestrial collider experiments, the forces acting on particles can be inferred from the trajectory and quantity of emerging material. 
Collisions between galaxy clusters, which contain dark matter, provide similar tests for dark sector forces. 
If dark matter's particle interactions are frequent but exchange little momentum (via a light mediator particle that produces a long-ranged force and anisotropic scattering), the dark matter will be decelerated by an additional drag force. % and become spatially offset. 
If the interactions are rare but exchange a lot of momentum (via a massive mediator that produces a short-ranged force and isotropic scattering), %corresponding to short-ranged forces via a massive mediator), 
dark matter will tend to be scattered away and lost \cite{impactpars,SIDMModel,bulletcluster}. 

The dynamics of colliding dark matter can be calibrated against that of accompanying standard model particles. 
The stars that reside within galaxies, which are visible in a smoothed map of their optical emission, have effectively zero cross-section because they are separated by such vast distances that they very rarely collide. 
The diffuse %mostly ionized %WHY IS IONISATION RELEVANT?
gas between galaxies, which is visible in X-ray emission, has a large electroweak cross-section; it is decelerated and most is eventually stripped away by ram pressure \cite{eckert14}. 
Dark matter, which can be located via gravitational lensing \cite{BS01}, behaves somewhere on this continuum (Fig.~1).

The tightest observational constraints on dark matter's interaction cross-section come from its behavior in the giant `bullet cluster' collision 1E0657-558 \cite{bulletclusterA}. 
A test for drag yields $\sdm<1.25\,\csg$ (68\% CL), and a test for mass loss yields $\sdm<0.7\,\csg$ (68\% CL) \cite{impactpars}. 
Half a dozen more galaxy cluster collisions have since been discovered, but no tighter constraints have been drawn. 
This is because the analysis of any individual system is fundamentally limited by uncertainty in the 3D collision geometry (the angle of the motion with respect to our line of sight, the impact parameter, and the impact velocity) or the original mass of the clusters. 

The same dynamical effects are also predicted by simulations in collisions between low-mass systems \cite{SIDMModel}.
Observations of low-mass systems produce noisier estimates of their mass and position \cite{cannibal, Harvey13,bulletgroup}, but galaxy clusters continually grow through ubiquitous `minor mergers', and statistical uncertainty can be decreased by building a potentially very large sample \cite{bulleticity,bulletgroupA}. 
Furthermore, we have developed a statistical model to measure dark matter drag from many noisy observations, within which the relative trajectories of galaxies, gas, and dark matter can be combined in a way that eliminates dependence upon 3D orientation and the time since the collision \cite{Harvey14}. 
%NEXT SENTENCE REMOVED; WHY NOT JUST WAIT UNTIL WE HAVE DONE THE CONTROL TEST (IN TWO PARAS) TO MENTION IT?
%This method also provides a control test. 

We have studied all galaxy clusters for which optical imaging exists in the Hubble Space Telescope (Advanced Camera for Surveys) data archive \cite{hst_archive_url} and X-ray imaging exists in the Chandra Observatory data archive \cite{chandra_archive_url}. 
We select only those clusters containing more than one component of spatially extended X-ray emission. 
Our search yields 30 systems, mostly between redshift $0.2<z<0.6$ plus two at $z>0.8$, containing 72 pieces of substructure in total (Table~S1). 
In every piece of substructure, we measure the distance from the galaxies to the gas $\sg$. 
Assuming this lag defines the direction of motion, we then measure the parallel $\si$ and perpendicular $\di$ distance from the galaxies to the lensing mass (Fig.~2).
%position of galaxies, gas and dark matter.
%{\bf We calculate the distance from the galaxies to the gas $\sg$, and from the galaxies to the dark matter, parallel $\si$ %and perpendicular $\di$ to the direction of motion (Figs.~1 \& 2).}
%VERY LONG SENTENCE!
%{\bf and calculate four distances: the amount the gas lags behind the galaxies, $\sg$, the perpendicular distance the dark matter centroid lies from the vector joining the gas and the galaxies, $\di$ (where the point where these two vectors meet is denoted the intersection point) and the distance from the intersection point (I) to the galaxies, $\si$, parallel to $\sg$, and finally the distance from the gas to the intersection point, $\gi$ (Fig.~1)}.  

We first test the null hypothesis that there is no dark matter in our sample of clusters (a similar experiment was first carried out on the Bullet Cluster, finding a 3.4 and 8$\sigma$ detection \cite{separation}).
 Observations that do not presuppose the existence of dark matter \cite{cosmosbaryon} show that $10^{14}M_\odot$ clusters contain only 3.2\% of their mass in the form of stars.
We compensate for this mass, which pulls the lensing signal towards the stars and raised $\gi$ by an amount typically $0.78\pm0.30$\,kpc (computed using the known distances to the stars $\sg$; see Materials and Methods).
The null hypothesis is that the remaining mass must be in the gas.
 However, we observe a spatial offset between that is far from the expected overlap, even in the presence of combined noise from our gravitational lensing and X-ray observations (Fig.~3A).
%Fig.~3a shows the expected overlap between total mass and gas, with the width of the distribution determined by the combined noise in our gravitational lensing and X-ray observations. The histogram in Fig.~3a shows the observed offset. 
A Kolmogorov-Smirnov test indicates that the observed offsets between gas and mass are inconsistent with the null hypothesis at 7.6$\sigma$, a p-value of $3\times10^{-14}$ (without compensation for the mass of stars, this is 7.7$\sigma$). This test thus provides direct evidence for a dominant component of matter in the clusters that is not accounted for by the luminous components. 

Having reaffirmed the existence of dark matter, we attempt to measure any additional drag force acting upon it, caused by long-range self-interactions. 
We measure the spatial offset of dark matter behind the stars, compensating as before for the 16\% of mass in the gas \cite{planckpars} by subtracting a small amount from $\si$ (on average $4.3\pm1.6$\,kpc).
%Fig.~3b shows the observed offsets $\si$ and $\sg$. 
%Similarly to before, we compensate for the 16\% of mass in the gas \cite{planckpars} by subtracting a small amount from $\si$ (on average $4.3\pm1.6$\,kpc). 
%It is difficult to compare offsets directly between collisions that have different physical scales. Nonetheless, w
We measure a mean dark matter lag of $\langle\si\rangle=-5.8\pm8.2$\,kpc in the direction of motion (Fig.~3B), and $\langle\di\rangle=1.8\pm7.0$\,kpc %1.77\pm6.96$\,kpc.
perpendicularly. 
The latter is useful as a control test: symmetry demands that it must be consistent with zero in the absence of systematics. We also use its scatter as one estimate of observational error in the other offsets.

%INCORPORATED INTO NEXT PARA
%For each system, we calculate the fractional lag of dark matter compared to gas, $\beta\equiv\si/\sg$.

%THIS PARAGRAPH LENGTHENED (USING PREVIOUS PARAGRAPH AND SOME SUPPLEMENTARY MATERIAL) TO BE EXPLICIT ABOUT WHAT WE'VE ASSUMED
We interpret the lag through a model \cite{Harvey14} of dark matter's optical depth (similarly to previous studies \cite{bulletcluster,cannibal}). 
Gravitational forces act to keep gas, dark matter and galaxies aligned, while any extra drag force on dark matter induce a fractional lag
\begin{equation}
\beta\equiv\frac{\si}{\sg}=B \left(1-\mathrm{e}^{\left(\frac{-(\sigma_{\rm DM}-\sigma_{\rm gal})}{\sstar}\right)} \right), \label{eqn:beta}
\end{equation}
 where $\sigma_{\rm gal}$ is the interaction cross-section of the galaxies, coefficient $B$ encodes the relative behavior of dark matter and gas, and $\sstar$ is the characteristic cross-section at which a halo of given geometry becomes optically thick. 
We assume that stars do not interact, so $\sigma_{\rm gal} \approx0$.
To ensure conservative limits on $\sdm$, we also assume $B\approx1$ and marginalize over $\sstar\approx6.5\pm3\,\csg$, propagating this broad uncertainty to our final constraints (see Materials and Methods).
%%The behaviour of dark matter is calibrated against that of gas, 
%and we conservatively assume $B\approx1$ (see Materials and Methods) and $\sstar$ is the characteristic cross-section at which a halo becomes optically thick.
%The behaviour of dark matter is compared to that of galaxies; since the gaps between stars are vast compared to their size, they interact very rarely, and we assume $\sigma_{\rm gal}\approx0$.
%For a $10^{14\pm 1}$ solar mass structure with typical geometry $\sstar\approx6.5\pm3\,\csg$; which we analytically marginalize over and propagate the uncertainty through to our final constraints.}
Adopting the dimensionless ratio $\beta$ brings two advantages. First, it removes dependence on the angle of the collision with respect to the line of sight. Second, it represents a physical quantity that is expected to be the same for every merger configuration, so measurements from the different systems can be 
%BEEN MORE EXPLICIT
%combined statistically
simply averaged (with appropriate noise weighting, although in practice, the constraining power from weak lensing-only measurements comes roughly equally from all the systems). 
%NEXT CLAUSE SUPPLIES NO MORE INFO, SO CUT TO SAVE SPACE
%, rather than being dominated by any single system. 

Combining measurements of all the colliding systems, we measure a fractional lag of dark matter relative to gas $\langle\beta\rangle=-0.04\pm0.07$ (68\% CL).
Interpreting this through our model implies that dark matter's momentum transfer cross-section is $\sdm=-0.25_{-0.43}^{+0.42}\,\csg$ (68\% CL, two-tailed), or $\sdm<0.47\,\csg$ (95\%CL, one-tailed); the full PDF is shown in Fig.~4. 
%This result thus disfavours any models of hidden dark sector physics that predict $\sdm\approx %1$\,barn/GeV$=
%0.6\,\csg$ with an anisotropic cross-section, e.g.\ \cite{mirror,glueballs,composite}. 
This result rules out parts of model space of hidden sector dark matter models e.g.\ \cite{mirror,glueballs,composite,resonant} that predict 
$\sdm\approx %1$\,barn/GeV$=
0.6\,\csg$ on cluster scales through a long-range force.
%ADDED THIS EXTRA SENTENCE; WE REFERRED TO THE CONTROL TEST ABOVE, SO IT'S NICE TO ACTUALLY QUOTE IT!
The control test found $\langle\beta_\perp\rangle\equiv\langle\di/\sg\rangle=-0.06\pm0.07$ (68\% CL), consistent with zero as expected.
This inherently statistical technique can be readily expanded to incorporate much larger samples from future all-sky surveys.
 Equivalent measurements of mass loss during collisions could also test dark sector models with isotropic scattering.
Combining observations, these \textit{astrophysically large particle colliders} have potential to measure dark matter's full differential scattering cross-section.

%MOVED TO SUPPLEMENTARY MATERIALS TO SAVE SPACE - EXTRA TESTS ARE FINE IN SUPPLEMENTARY
%Note that, if we repeat the analysis measuring mean galaxy positions from the number-weighted distribution of galaxies (rather than the flux-weighted distribution -- so at the other extreme, and less realistic assumption of mass/light ratio), we obtain consistent values of $\langle\beta\rangle=0.054\pm0.062$ (68\% CL) and conclude that $\sdm=0.36_{-0.45}^{+0.46}\,\csg$ (68\% CL, two-tailed).

\clearpage
\begin{figure}[t]
\centering
\includegraphics[width=0.8\textwidth, trim = 77mm 118mm 77mm 110mm]{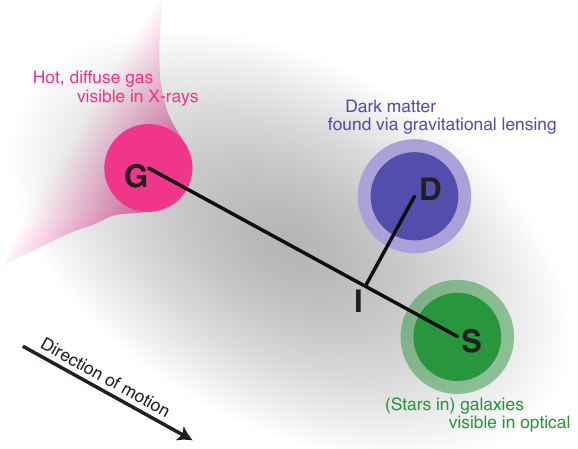}
\caption{Cartoon showing the three components in each piece of substructure, and their relative offsets, illustrated by black lines. The three components remain within a common gravitational potential, but their centroids become offset due to the different forces acting on them, plus measurement noise. We assume the direction of motion to be defined by the vector from the diffuse, mainly hydrogen gas (which is stripped by ram pressure) to the galaxies (for which interaction is a rare event).  We then measure the lag from the galaxies to the gas $\sg$, and to the dark matter in a parallel $\si$ and perpendicular $\di$ direction. %We then measure offsets to the intersection point, I, defined by extending a perpendicular line from the location of the dark matter.
}
\label{fig:cartoon}
\end{figure}

\clearpage
\begin{figure}[t]
\centering
\includegraphics[width=\textwidth, trim = 15mm 40mm 15mm 20mm]{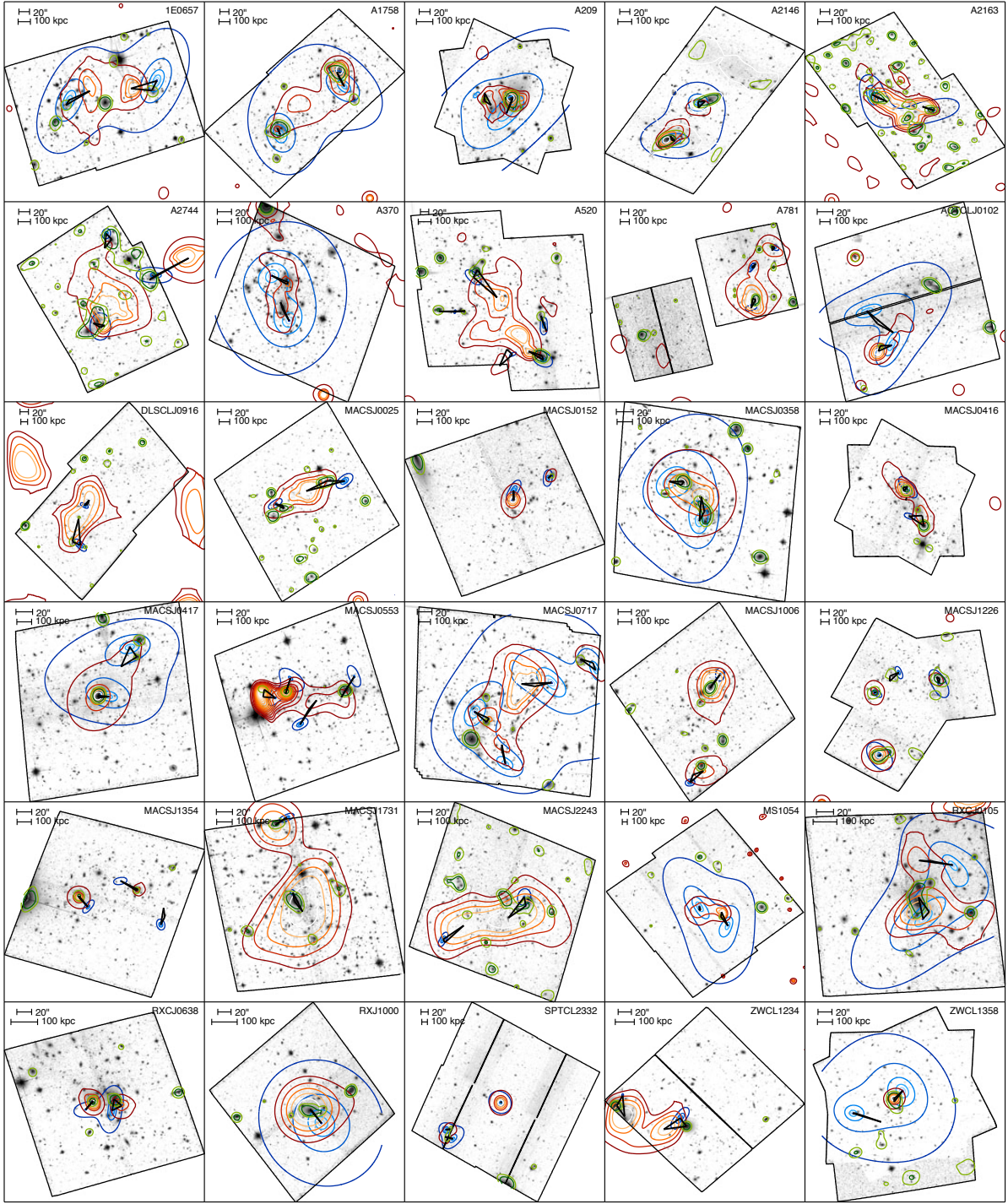}
\caption{Observed configurations of the three components in the 30 systems studied. The background shows the HST image, with contours showing the distribution of galaxies (green), gas (red) and total mass, which is dominated by dark matter (blue).}
\label{fig:observations}
\end{figure}

\clearpage
\begin{figure}[t]
\centering
\includegraphics[width=0.8\textwidth, trim = 77mm 112mm 77mm 110mm]{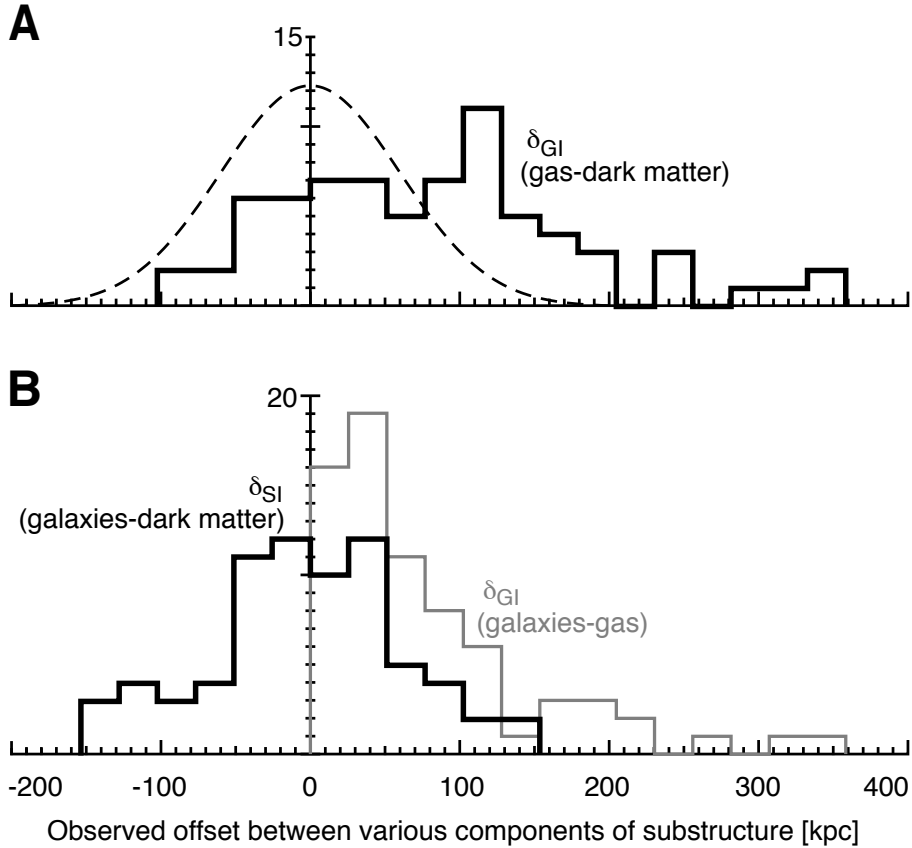}
\caption{Observed offsets between the three components of 72 pieces of substructure. Offsets $\si$ and $\gi$ include corrections accounting for the fact that gravitational lensing measures the total mass, not just that of dark matter. (A) The observed offset between gas and mass, in the direction of motion. The smooth curve shows the distribution expected if dark matter does not exist; this hypothesis is inconsistent with the data at 7.6$\sigma$ statistical significance. (B) Observed offsets from galaxies to other components. The fractional offset of dark matter towards the gas, $\si/\sg$, is used to measure the drag force acting on the dark matter.}
\label{fig:offsets}
\end{figure}

\clearpage
\begin{figure}[t]
\centering
\includegraphics[width=0.8\textwidth, trim = 77mm 122mm 77mm 124mm]{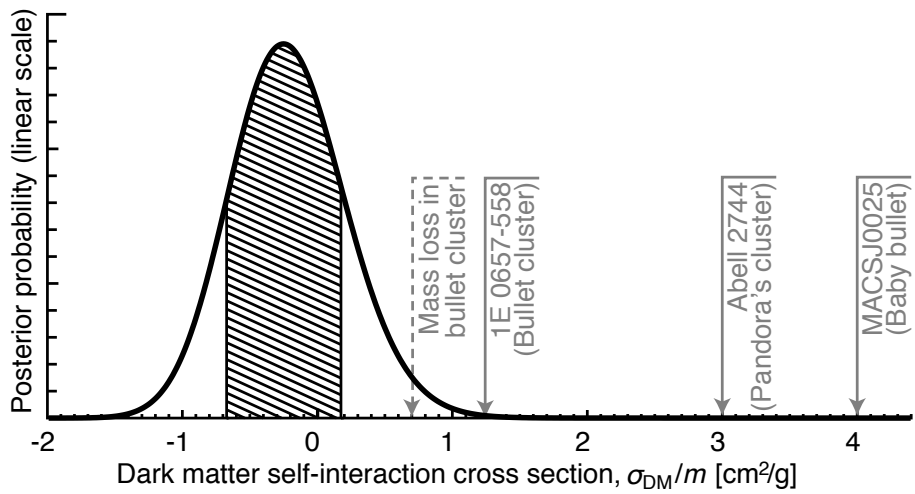}
\caption{Constraints on the self-interaction cross-section of dark matter. These are derived from the separations $\beta=\si/\sg$, assuming a dynamical model to compare the forces acting on dark matter and standard model particles \cite{Harvey14}. The hatched region denotes 68\% confidence limits, to be compared to the 68\% confidence upper limits from previous studies of the most constraining individual clusters in blue. Note that the tightest previous constraint is derived from a measurement of dark matter mass loss, which is sensitive to short range self-interaction forces; all other constraints are measurements of a drag force acting on dark matter, caused by long range self-interactions.}
\label{fig:constraints}
\end{figure}
\clearpage

\subsection*{Acknowledgements}
DH is supported by the Swiss National Science Foundation (SNSF) and STFC. 
RM and TK are supported by the Royal Society.
The raw HST and Chandra data are all publicly accessible from the mission archives \cite{hst_archive_url,chandra_archive_url}.
We thank the anonymous referees, plus Scott Kay, Erwin Lau, Daisuke Nagai and Simon Pike for sharing mock data on which we developed our analysis methods; Rebecca Bowler for help stacking HST exposures; Eric Jullo, Jason Rhodes and Phil Marshall for help with shear measurement and mass reconstruction; Doug Clowe, Hakon Dahle and James Jee for discussions of individual systems; Celine Boehm, Felix Kahlhoefer and Andrew Robertson for interpreting particle physics.

\newpage

\setcounter{figure}{0} \renewcommand{\thefigure}{S\arabic{figure}} 
\setcounter{equation}{0} \renewcommand{\theequation}{S\arabic{equation}} 
\setcounter{page}{1}

\begin{figure}[t]
\centering
\includegraphics[width=0.2\textwidth]{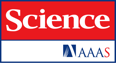} \vspace{10mm}
\end{figure}

\section*{Supplementary materials for\\~\\ \papertitle}

\noindent David Harvey$^{1,2}$, Richard Massey$^{3}$, Thomas Kitching$^{4}$, Andy Taylor$^2$ \& Eric Tittley$^2$\\
~\\
\small{$^{1}$Laboratoire d'astrophysique, EPFL, Observatoire de Sauverny, 1290 Versoix, Switzerland}\\
\small{$^{2}$Royal Observatory, University of Edinburgh, Blackford Hill, Edinburgh EH9 3HJ, UK}\\
\small{$^{3}$Institute for Computational Cosmology, Durham University, South Road, Durham DH1 3LE, UK}\\
\small{$^{4}$Mullard Space Science Laboratory, University College London, %Holmbury St Mary, 
Dorking, Surrey RH5 6NT, UK}\\
~\\
\small{Correspondence to: david.harvey@epfl.ch}

\vspace{20mm}
{\bf This PDF file includes:}
\begin{itemize}
\item Materials and Methods
\item SupplementaryText
\item Figs. S1 to S8
\item References 32--47
\end{itemize}

\newpage

\section*{Materials and methods}

We followed overall procedures that we developed in blind tests on mock data \cite{Harvey13}, usually exploiting algorithms for high precision measurement that had been developed, calibrated and verified elsewhere.
However, several custom adaptations were required to analyze the heterogeneous data from the Hubble Space Telescope (HST) and Chandra X-ray Observatory archives (Table S1 lists all the observed systems \cite{MACS_mergers,MACS_brightest,MACS_MACS,MACS_sample}, and Figure S2 shows the offsets measured in each).

Here we describe the methods we used to combine observations with different exposure times, filters, epochs and orientations -- starting from the raw data and performing a full reduction to maximise data quality. To convert angular distances into physical distances, we assume a cosmological model derived from measurements of the Cosmic Microwave Background \cite{planckpars}, $\Omega_{\rm M}=0.31$, $\Omega_{\rm \Lambda}=0.69$, $H_0=67$\,km/s/Mpc.

\begin{figure}
\centering
\includegraphics[width=0.9\textwidth]{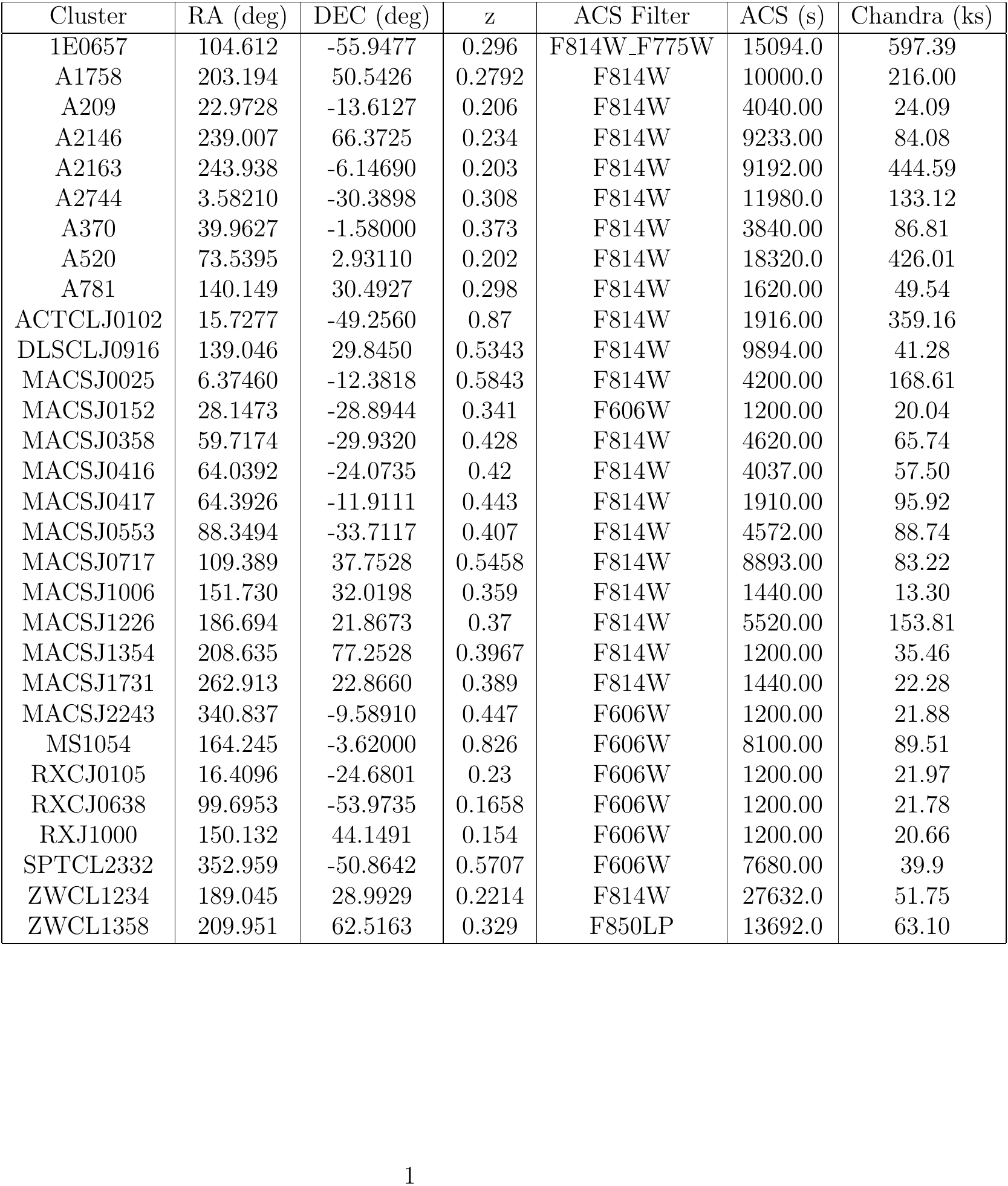}
\caption{The full sample of 30 merging complexes, and their locations on the sky. The columns show, from left to right: the name of the cluster, its right ascension, declination, and redshift, the HST/ACS filter used for our lensing analysis, and the total exposure time for that particular filter, and the (cleaned) exposure time of the Chandra X-ray image.}
\label{fig:sample}
\end{figure}

\begin{figure}
\centering
\includegraphics[height=0.85\textheight]{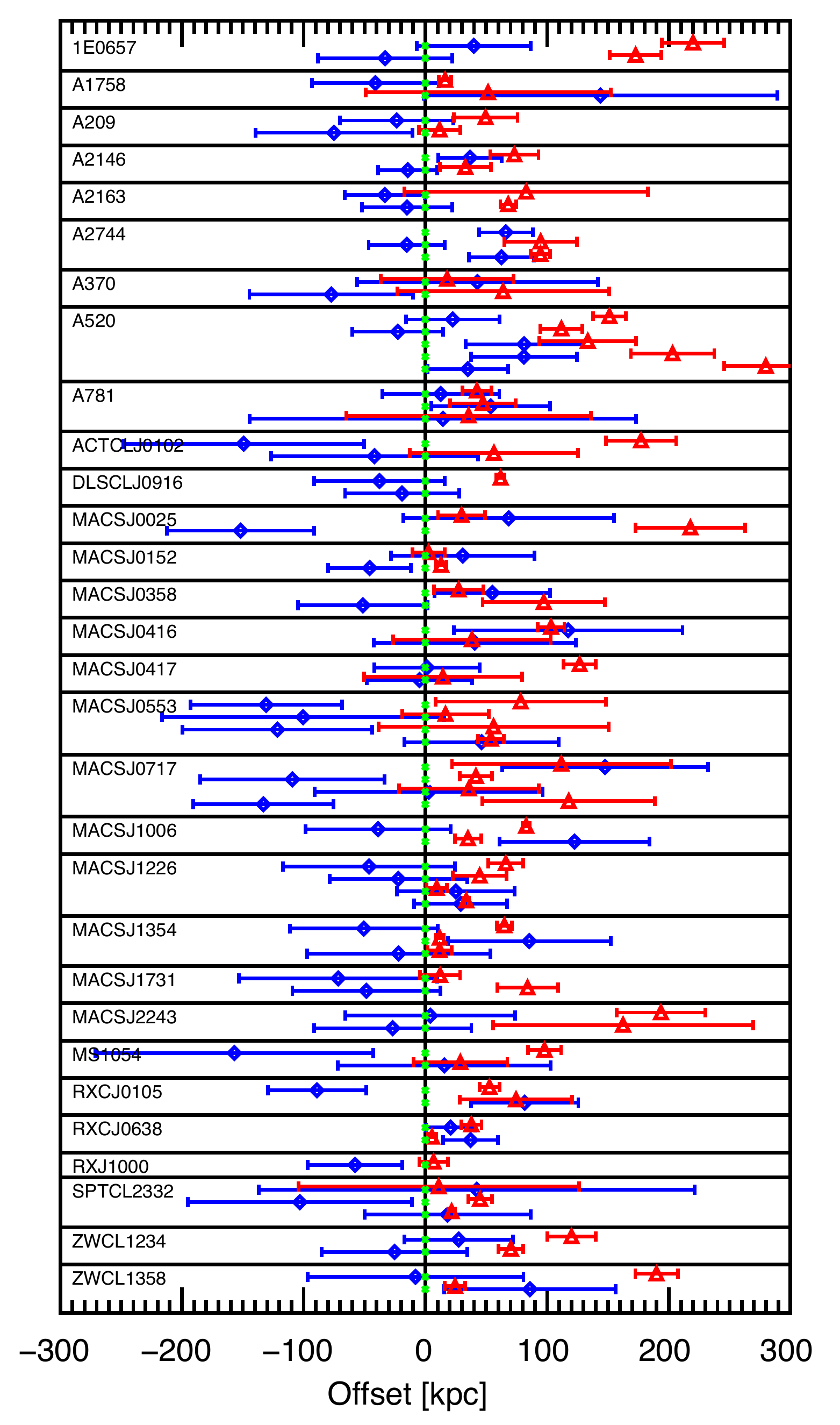}
\caption{Observed offsets between galaxies, gas and dark matter in 72 components of substructure. In each case, the green triangle, at the centre of the coordinate system, denotes the position of the galaxies. The separation between galaxies and gas, $\sg$, is shown in red. The separation of the dark matter with respect to the galaxies, projected onto the SG vector, $\si$, is shown in blue. The error bars show the locally estimated $1\sigma$ errors.}
\label{fig:offsetstable}
\end{figure}

\subsection*{Position of gas, seen in X-ray emission}

We downloaded the raw~event~1 files for all observations. To process these data, we used CIAO tools version~4.5, starting with basic reduction and calibration using the CIAO {\it repro} tool. In our analysis, it is particularly important to remove emission from point sources, and prevent X-ray bright Active Galactic Nuclei at the centers of clusters from biasing our position measurements. We therefore made a first pass at removing point sources using {\it celldetect}. We then filtered each event table for any potential spurious events such as solar flares by clipping the table at the $4\sigma$ below the mean flux level. 

%Having cleaned each exposure, we combined them using the {\it merge\_obs} script from CIAO tools into a single exposure map corrected flux image, producing along with it exposure maps for each observation and the stacked image. To smooth the image, we use {\it wavdetect} CIAO, a wavelet smoothing algorithm that employs a `Mexican hat' filter on a range of scales. This uses the given size of the Chandra PSF at each position in the field to calculate the estimated size of a source; to model the PSF throughout the stacked image we created individual maps using {\it mkpsfmap} for each exposure at an effective energy of 1\,keV, then combined each model weighting them by their respective exposure map. Figure~S3 shows an example of the PSF map used for cluster A520. 

Having cleaned each exposure, we combined them using the {\it merge\_obs} script from CIAO tools into a single exposure map corrected flux image, producing along with it exposure maps for each observation and the stacked image. 
We modeled the Chandra PSF at each position throughout the field, we created individual maps using {\it mkpsfmap} for each exposure at an effective energy of 1\,keV, then combined each model weighting them by their respective exposure map. Figure~S3 shows an example of the PSF map used for cluster A520. 

To make a second pass to identify point sources, we passed the stacked image and PSF model through CIAO {\it wavdetect}, a wavelet smoothing algorithm that employs a `Mexican hat' filter on a range of scales. This estimates the true size of each source, correcting for the size of the PSF. We used the smallest scales for the wavelet radii (1, 2\,pixels) to identify point sources, and combined the larger scales (4, 8, 16, and 32\,pixels) into a denoised version of the final image. We finally inspected every image by eye for any remaining point sources. We found that this double filter method proved very successful at removing point sources, with only AGN at the edge of the cluster remaining unflagged. Although their emission has extended wings, the cluster is usually in the center of the pointing, resulting in minimal contamination. 

\begin{figure}
\centering
\includegraphics[width=0.6\textwidth]{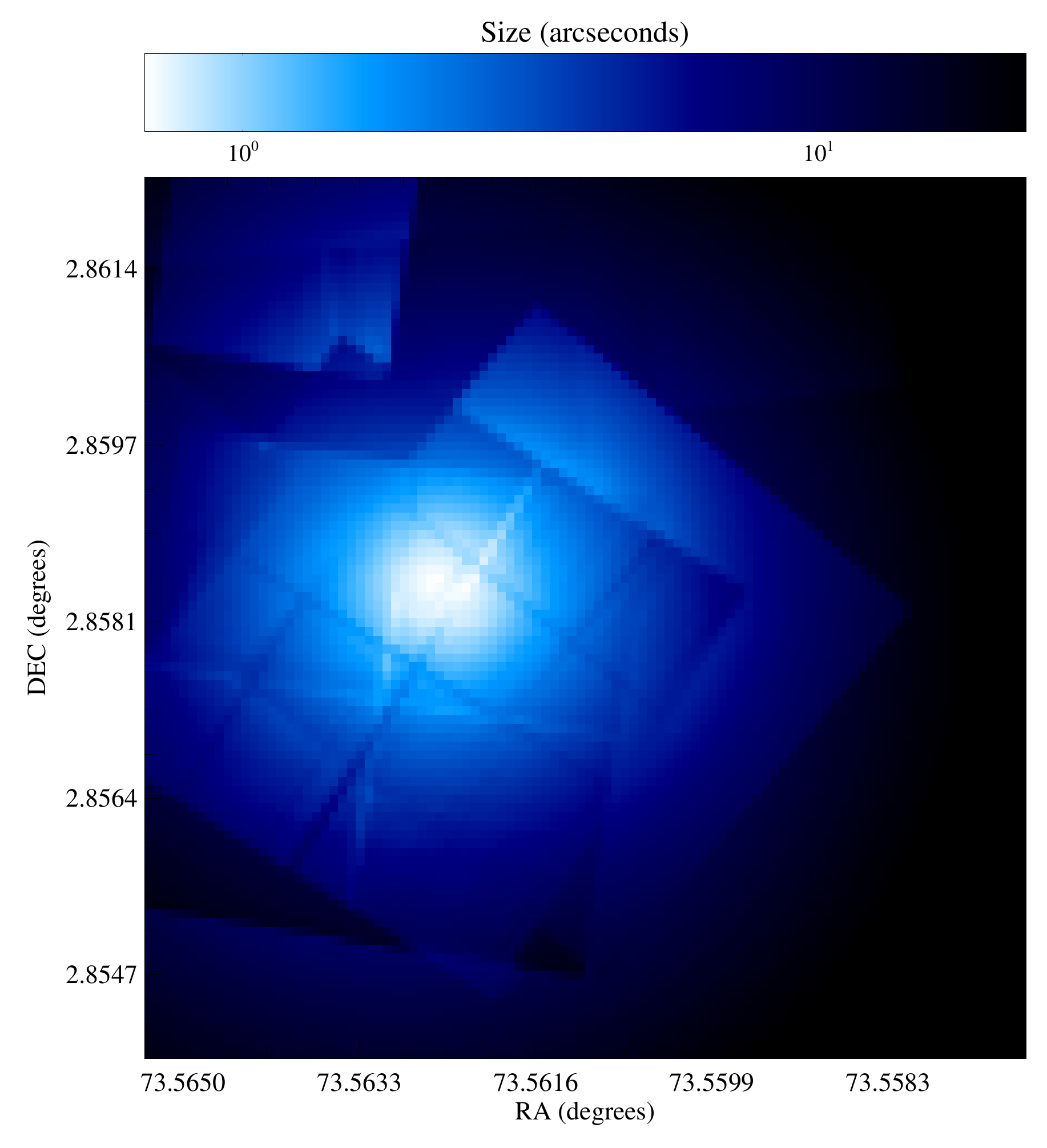}
\caption{An example model of the size of the Chandra X-ray telescope's Point Spread Function (PSF). The model PSF is used to identify and remove point sources, e.g.\ Active Galactic Nuclei -- and to thereby identify extended X-ray emission from hot gas within the cluster. The image shows a combined, exposure map weighted, PSF map stacked for the various observations of galaxy cluster A520.}
\label{fig:chandrapsf}
\end{figure}

%In order to smooth the data, we first passed the stacked image and PSF model through {\it wavdetect} using the smallest scales for the wavelet radii (1, 2\,pixels) as a secondary filter for point sources and then again using scales 4, 8, 16, and 32\,pixels to obtain the final image. We further inspected every image by eye for any remaining point sources. We found that this double filter method proved very successful at removing point sources, with only AGN at the edge of the cluster remaining unflagged. Although their emission has extended wings, the cluster is usually in the center of the pointing, resulting in minimal contamination. 

Finally, we measured the position of coherent substructure in the X-ray emission using {\it SExtractor} \cite{sextractor}. This calculates positions from the first order moments of the light profile, which means that the returned position does not always coincide exactly with the brightest pixel. {\it SExtractor} does not report reliable errors in the positions but, since the dominant contribution of variation is the size of the smoothing kernel, we can estimate the robustness of our measurements by smoothing the image using different scales in {\it wavdetect}, and measure the rms across different scales. On average we found the rms error to be 4\,arcseconds (roughly 30\,kpc at redshift $z$=0.4).

\subsection*{Position of galaxies, seen in optical emission}

We searched the HST archive for data acquired with the Advanced Camera for Surveys (ACS) instrument, which has the largest field of view. We considered only filters F606W, F814W and F850LP, whose high throughput ensures deep imagining, and whose red wavelengths ensure (a) that the optical emission samples the old stars that dominate the mass content of these systems and (b) a high density of high redshift galaxies visible behind the cluster, to provide sufficient lensing signal. Some clusters had been observed in more than one wavelength band. We used only a single band for all the clusters to further homogenize the data, but have compared a subset of our results in different bands to check for systematic errors. For our main analysis, we selected the broad F814W band, unless there are significantly more exposures in another. %; table S1 lists the exposure time in the selected filter.

We corrected the raw, pixellated data for charge transfer inefficiency \cite{CTI2}, then performed basic data reduction and calibration using the standard {\it Calacs} pipeline. We used {\it tweakReg} to orient and align individual exposures, then stacked them using {\it MultiDrizzle} \cite{astrodrizzle} with a Gaussian convolution kernel and {\sc pixfrac=0.8} \cite{drizzlepars} to produce a deep, mosaicked image with a pixel scale of 0.03\,arcseconds. In the process, {\it MultiDrizzle} also output a reoriented image of each individual exposure, which we used for star/galaxy identification and PSF estimation.

We estimated the distribution of mass in galaxies via the proxy of the light emitted by their stars. In our single-band imaging, we were able to identify and mask foreground stars in the Milky Way (which appear pointlike), but assumed any foreground or background galaxies to be randomly positioned and thus merely add shot noise to our measurements. We smoothed the masked image using {\it wavdetect}, and measured the position of coherent substructure using {\it SExtractor} \cite{sextractor}. This calculates positions from the first order moments of the light profile, which means that the returned position does not always coincide exactly with the brightest pixel. {\it SExtractor} does not report reliable errors in the positions. However, since the dominant contribution of noise is inclusion or omission of galaxies inside the smoothing kernel, we estimated the robustness of our measurements by smoothing the image using different scales in {\it wavdetect}, and compared the resulting positions. On average, we found an rms error in the position of the extracted halos of 0.6\,arcseconds (roughly 4.5\,kpc at redshift $z$=0.4).

We also tried two other ways to quantify the position of the galaxies. First, we measured the smoothed distribution of galaxies in the image, with all galaxies weighted equally (this represents the opposite -- and least realistic -- assumption of galaxies' mass/light ratio). To do this in practice, we passed the galaxy catalogue through the X-ray data reduction pipeline, as if each galaxy were a single X-ray photon. This created a smoothed image, in which we identified substructure using {\it SExtractor}. Since the same galaxies contributed both to the flux-weighted and galaxy-weighted positions, the two measurements are correlated. We measure the uncertainty on the galaxy weighted positions to be 5\,kpc, about the same as the flux-weighted positions. 
%MOVED HERE FROM MAIN TEXT
%Note that, if we repeat the analysis measuring mean galaxy positions from the number-weighted distribution of galaxies (rather than the flux-weighted distribution -- so at the other extreme, and less realistic assumption of mass/light ratio), 
We obtain consistent values of $\langle\beta\rangle=0.054\pm0.062$ (68\% CL) and conclude that $\sdm=0.36_{-0.45}^{+0.46}\,\csg$ (68\% CL, two-tailed).
Second, we tried identifying the position of the `Brightest Group Galaxy' (BGG), since its formal error is small, and it has proved optimal in studies of isolated groups \cite{galgroup}. In merging systems however, the brightest nearby galaxy is frequently unassociated with the infalling group \cite{BCGmerger}. Accounting for our observed $1.7\pm0.9$\,arcsecond offset to any brighter galaxy within 25\,arcseconds of X-ray emission (the search region that will be used to identify gravitational lensing signals), again yields a consistent constraint on $\sdm$, but with much larger final error.

\subsection*{Position of dark matter, measured via weak gravitational lensing}

We measured the ellipticities of galaxies in HST images using the {\it RRG} method \cite{RRG}. This corrects galaxies' Gaussian-weighted moments for convolution with the Point Spread Function (PSF), to measure the shear $\gamma_1$ ($\gamma_2$) corresponding to elongations along (at 45 degrees to) the $x$ axis. This method has been empirically calibrated on simulated HST imaging in which the true shear is known \cite{COSMOSintdisp}, applying a multiplicative correction of $\langle m\rangle=-3.0\times10^{-3}$ and a additive bias of $\langle c\rangle=-2.1\times10^{-4}$.

HST's PSF varies across the field of view and, because thermal variations change the telescope's focus, at different epochs. Modelling the net PSF in our stacked images therefore required a flexible procedure. We first identified stars in the deep, stacked image using their locus in size--magnitude space. We then measured the ellipticity of each star in individual exposures. By comparing these to {\it TinyTim} \cite{tinytim} models of the HST PSF (created by raytracing through the telescope at different focus positions but at the appropriate wavelengths for the band), we determined the focus position for each exposure. We then interpolated (second and fourth shape moments of) the {\it TinyTim} PSF model to the position of the galaxies, rotating into the reference frame of the {\it MultiDrizzle} mosaic. We then summed the PSF moments from each exposure in which a galaxy was observed. Figure~S4 shows an example of the final PSF model for one cluster.

\begin{figure}
\centering
\includegraphics[width=0.7\textwidth]{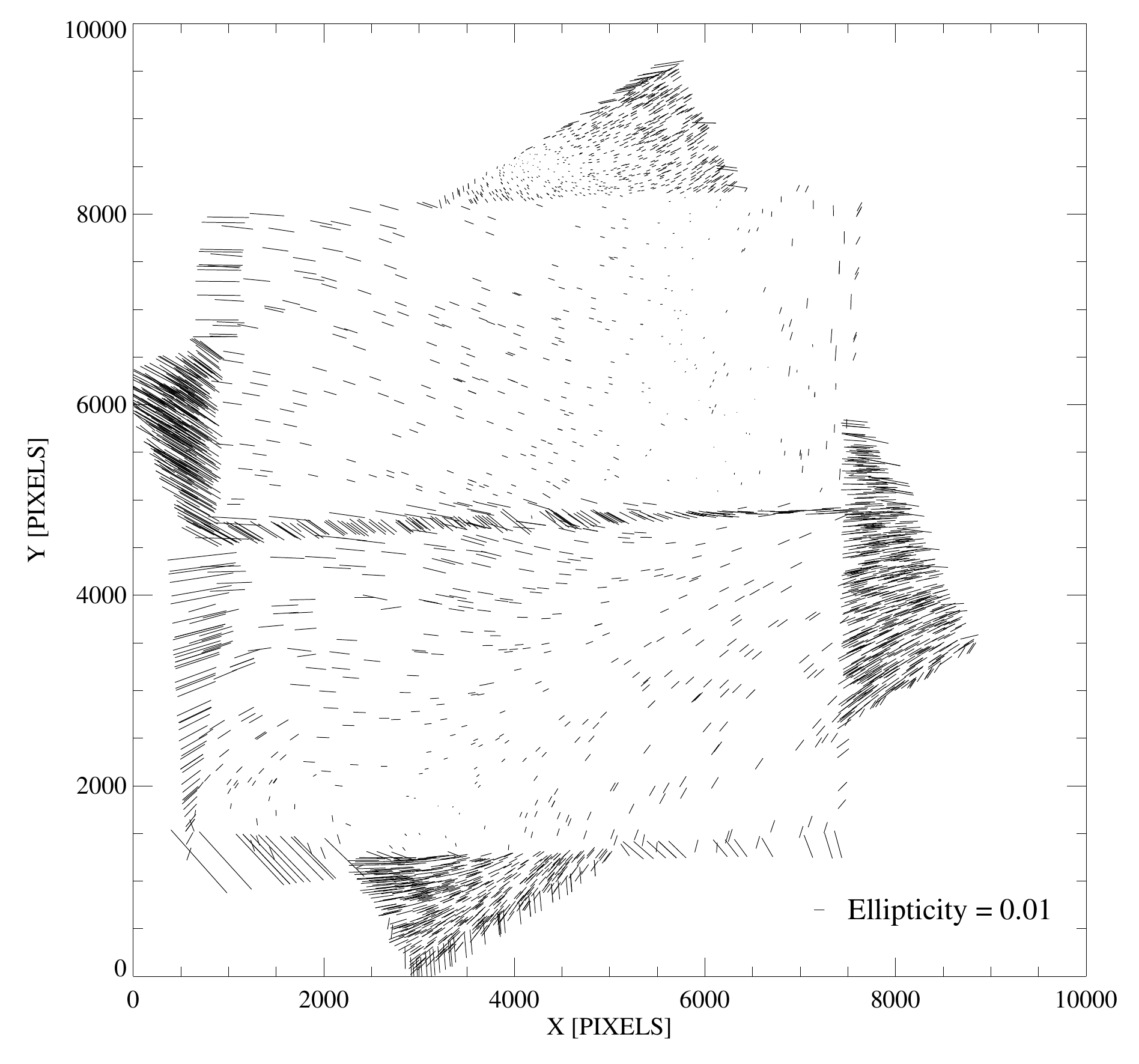}
\caption{An example model of the Point Spread Function (PSF) of the Hubble Space Telescope/Advanced Camera for Surveys (HST/ACS). Each tick mark represents the ellipticity of the PSF at that particular position in the HST field. Its orientation shows the PSF's major axis and its length shows the ellipticity; a dot would indicate a circular PSF. The PSF tends to be highly elliptical near the edge of the field and more circular in the centre. Tick marks are plotted at the position of every ``detected" source. The mosaic pattern of dithered exposures can be seen: noisier regions with fewer exposures contain more spurious sources, which are removed during analysis (but are shown here for clarity). The example shown is for observations of galaxy cluster MACSJ0416.}
\label{fig:hstpsf}
\end{figure}

We measured the shear of all galaxies that appear in $3$ or more exposures, with a combined signal-to-noise in the stacked image $>4.4$ and size $>0.1$\,arcseconds. These cuts \cite{COSMOSintdisp} remove noisy measurements at the edges of the field or in the gaps between detectors. We also masked out galaxies that lie near bright stars or large galaxies, whose shapes appear biased. Figure~S5 compares shear catalogues for a single cluster, derived from independent analyses of data in the F814W and F606W bands. There is the expected level of scatter between the two measurements -- but, most importantly, there is no detectable bias.

\begin{figure}
\centering
\includegraphics[width=0.7\textwidth]{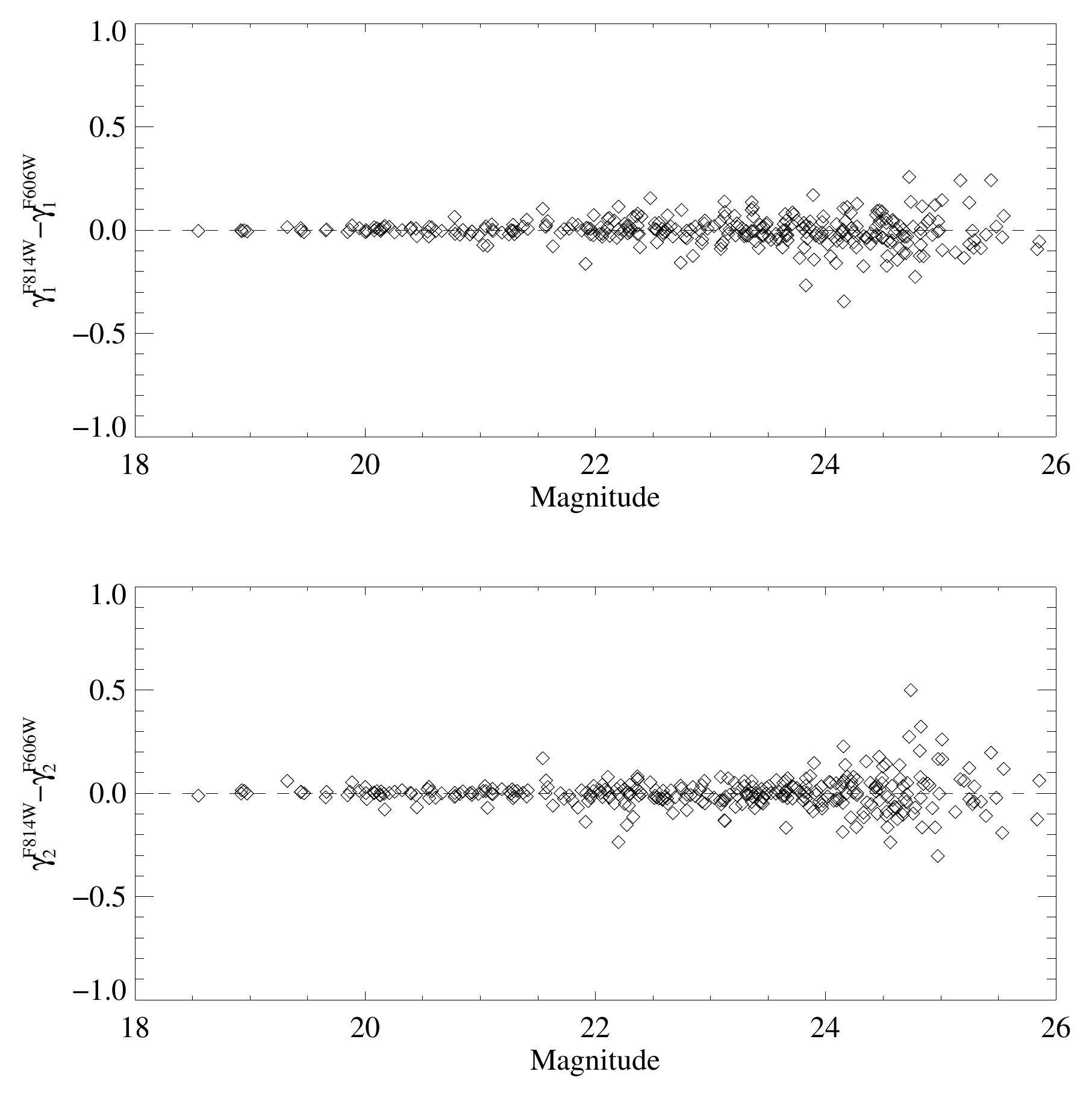}
\caption{A comparison of the gravitational lensing shears measured independently behind a single cluster, in two different HST filters. The top (bottom) panel shows the difference between $\gamma_1$ ($\gamma_2$) for each galaxy, which traces to elongations along (at 45 degrees to) the $x$ axis. We find scatter as expected due to observational noise, but no systematic bias.}
\label{fig:2bands}
\end{figure}

We reconstructed the distribution of mass in the clusters using the parametric model-fitting algorithm {\it Lenstool} \cite{lenstool}. Using Bayesian likelihood minimization, {\it Lenstool} simultaneously fits multiple mass haloes to an observed shear field, with the position and shape of each halo described by the NFW \cite{NFW} density profile. This is an efficient technique to record a unique position for each halo, marginalizing over nuisance parameters that  include mass and morphology, that are not of direct interest to our study. Assuming this density profile does not bias measurements of the position of halos within current statistical limits \cite{Harvey13}. 
{\it Lenstool} requires positional priors to be defined in which it searches for the lensing signal. Except in a few well-studied systems (where we use the extra information), we obtained an initial lensing model using one prior search radius centered on each gas position and large enough to incorporate any nearby groups of galaxies. Following this scheme, we used an automated procedure to identify and associate the mutually closest galaxy, gas and lensing signals into systems of three mass components. In all systems, we then modeled the lensing system a final time, adopting priors centred on the galaxy position (we redid this step when trying different position estimators for the galaxies). Henceforth, we could center the coordinate system for each combined system of galaxies, gas and dark matter on the galaxies, to avoid prior bias in the Bayesian fits.

{\it Lenstool} samples the posterior surface in two ways. To obtain the best fitting position, we iterated to the best-fit solution with a converging MCMC step size, using ten simultaneous sampling chains to avoid local maxima. To sample the entire posterior surface (whose width quantifies uncertainty on model parameters), we then reran the algorithm with a fixed step size. The $1\sigma$ error on position was on average 11.4\,arcseconds (roughly 60\,kpc at redshift $z$=0.4). {As a sanity check we compare our measured centroids to those systems included in previous studies. Our statistical uncertainty is sometimes larger because we use only weak gravitational lensing, but we find no evidence for any bias. For example, our measured positions in the `bullet cluster' lie within one standard deviation of those reported in \cite{separation}}.

\subsection*{Positional offsets between components}

When assigning different mass components to one another, for almost all the clusters, we used an automated matching algorithm to associate the nearest clumps of dark matter, gas and stars. This was made robust by performing the matching in both directions (e.g., dark matter to stars, and stars to dark matter). In a few cases where detailed analyses of individual systems were available in the literature (for example, using strong lensing, X-ray shocks, optical spectroscopy or imaging additional bands, which were outside the scope of our work), we inserted that prior information by hand during association. This was most useful in systems A520 and A2744. { As a further test, we carry out a jackknife test to ensure that the association does not effect the overall constraints, and moreover, no single cluster dominates the result. We find no evidence for such an effect, and derive consistent error bars of $\Delta\sigma_{\rm DM, JK}/m=\pm0.5\csg$, further supporting the error bars quoted in our final result}. 

We drew an offset vector $\sg$ in angle between the observed position of the gas and galaxies, which we took to define the system's direction of motion. We then measured the position of the total mass along that vector and (in a right handed coordinate system) perpendicular to it, defining offset vectors $\si$, $\gi$, and $\di$ from the intersection point I of these vectors. 

Gravitational lensing measures the position of total mass, rather than that of just dark matter. We corrected the measured offsets $\si$ and $\gi$ for the contribution from the next most massive component. To calibrate this correction, we analysed mock lensing data from a dominant mass component (with an NFW \cite{NFW} profile) plus a less massive component at some offset $\delta$. The corrections were always small but, for a subdominant component with the same profile, normalised to contain a fraction $f$ of the total mass, we found that the lensing position is pulled by an amount $f\delta e^{-0.01\delta/r_s}$, and we corrected for that. {If we do not calibrate for the extra pull of gas on the lensing peak we infer an upper limit of $\sdm < 0.54\csg$ (68\% CL, one-tailed).}

To test the hypothesis that dark matter does not exist, we required a model of the $\gi$ data expected if this were true. To generate that model, we assumed that the true positions of the X-ray and lensing signals coincided, but that the observed positions were offset by a random amount determined by the appropriate level of noise in each (see above). We calculated the 2D offset, then projected this onto the direction to the stars, which is also selected at random. We could have slightly increased the model $\gi$ offset to account for the mass in stars (the increase must be positive because the vector $\sg$ is defined from the galaxies to the gas). However, it is better to instead decrease the observed $\gi$ offset. The two approaches are equivalent in principle, but the latter allowed information to be added to our analysis because the absolute value of $\sg$ was known in each system. When comparing the model and observed $\gi$ offsets via a Kolmogorov-Smirnov test (in which we computed critical values using a Taylor series), we also used the errors on $\sigma_{\rm GI}$ determined for each system individually.

When measuring the interaction cross-section of dark matter, we converted offset measurements in arcseconds to physical units of kpc (using a standard cosmological model, which assumes dark matter exists). This enabled a more detailed comparison of the offsets between different systems. The (noisily determined) error estimates of offsets in a few systems were anomalously low, and likely smaller than the uncertainty in our knowledge of the merging configuration. To more robustly quantify the total uncertainty of offsets (which should include observational noise plus the possibility of component misidentification and merging irregularities), we empirically exploited the control test $\di$, which has an rms variation between systems $\sigma_{\rm DI}=60$\,kpc. This value is consistent with most of the individually measured errors, but more robust. We therefore adopted it globally as the error on every measurement of $\di$ and $\si$, rescaling to a value in arcseconds at the redshift of each system. Errors in $\sg$ must be smaller than this, because they do not involve observational noise in the lensing position. However, they also include the possibility of component misidentification, which is best estimated through this global approach. We therefore adopted the conservative approach of also assigning this value as the error on every measurement of $\sg$. Thus we set $\sigma_{\rm SG}=\sigma_{\rm SI}=\sigma_{\rm DI}=60$\,kpc. To combine our measurements of $\beta=\si/\sg$ and $\beta_\perp=\di/\sg$ from individual systems, we multiplied their posterior probabilities (approximated as a normal distribution even though it is a Cauchy distribution, but with a width determined by propagating errors on the individual offsets). 
%MOVED TO MAIN TEXT
%We found $\langle\beta_\perp\rangle=-0.059\pm0.066$ (68\% CL).

\subsection*{Interpreting positional offsets as an interaction cross-section}

Similarly to previous studies of the cross-section of dark matter \cite{bulletcluster,cannibal}, we interpreted observations of offset dark matter in terms of its optical depth for interactions. However, {we have developed a more sophisticated model \cite{Harvey14} intended to take into account the 3D and time-varying trajectories of infalling halos. 
First, calculating the dimensionless ratio $\beta=\si/\sg$ removes dependence on the angle of the collision with respect to the line of sight. Furthermore, a set of analytic assumptions suggests that $\beta$ is a physically meaningful quantity that should be the same for every system.
The main assumption of quasi-steady state equilibrium is reasonable for the detectable systems in our sample, but caution would be needed to interpret dark matter substructure that had passed directly through the cluster core (and had its gas stripped) or substructure on a radial orbit caught at the brief moment of turnaround (this is a negligible fraction in our mock data).
The model also incorporates the results of simulations \cite{SIDMModel} in which dark sector interactions that are frequent but exchange little momentum (e.g.\ via a light mediator particle that produces a long-ranged force and anisotropic scattering) produce a drag force and separate dark matter from the stars. 
On the other hand, simulations of `billiard ball' interactions that are rare but exchange a lot of momentum (e.g.\ via a massive mediator that produces a short-ranged force and isotropic scattering) %corresponding to short-ranged forces via a massive mediator), 
tend to scatter dark matter away from a system and produce mass loss \cite{impactpars,SIDMModel,bulletcluster}. 
However, we note that the ref.\ \cite{impactpars} also reports an unexpected small separation between galaxies and dark matter after billiard ball scattering.
In this paper, we explicitly follow the prescription in \cite{SIDMModel}.

According to our model of dark matter dynamics (see equation 33 of ref.\ \cite{Harvey14}), the offset of dark matter from galaxies, calibrated against the offset of gas, is
%\begin{equation}
%\langle\beta\rangle=\frac{C_{\rm DM} A_{\rm DM} \rho^\mathrm{ICM}_{\rm DM}M_{\rm gas}}{C_{\rm gas}A_{\rm gas} \rho^\mathrm{ICM}_{\rm gas}M_{\rm DM}}\left(1-\mathrm{exp}{\left(\frac{-(\sdm-\sigma_{\rm gal})}{\sstar}\right)} \right).
%\end{equation}
%\begin{equation}
%\langle\beta\rangle=B\left(1-\mathrm{exp}{\left(\frac{-(\sdm-\sigma_{\rm gal})}{\sstar}\right)} \right),\mathrm{~where~}B=\frac{C_{\rm DM} A_{\rm DM} \rho^\mathrm{ICM}_{\rm DM}M_{\rm gas}}{C_{\rm gas}A_{\rm gas} \rho^\mathrm{ICM}_{\rm gas}M_{\rm DM}}. %=\mathcal{O}(1).
%\end{equation}
\begin{equation}
\langle\beta\rangle=B\left(1-\mathrm{exp}{\left(\frac{-(\sigma_{\rm DM}-\sigma_{\rm gal})}{\sigma^\star}\right)} \right).
\end{equation}
Since the gaps between galaxies are vast compared to their size, they interact very rarely, so we assumed that $\sigma_{\rm gal}\approx0$. If this assumption were wrong, or in the presence of observational noise, our analysis can therefore produce negative values of $\sdm$. 
Our quoted errors include observational errors, pair-assignment errors, and model parameter errors.

The value of $\sstar$ depends upon the geometrical properties of the dark matter halo, but is proportional to its mass and inversely proportional to its cross-sectional area. For our set of merging systems, we conservatively adopted a $\sstar=6.5\pm3\,\csg$ by assuming the system masses are $\log(M_{200}/M_\odot)=14\pm1$, with NFW density profiles and concentration varying with mass as observed in numerical simulations \cite{maccio}. {By assuming a conservative range in halo masses we propagate a much larger error in $\sstar$ than one would expect if we were to measure the true values.} We then analytically marginalized over $\sstar$, propagating the uncertainty through to our final constraints. The top panel of Figure~S6 shows the values of $\sstar$ instead assuming different, fixed system masses; the bottom panel shows the effect on $\sdm$. The inferred estimate of $\sdm$ is broadly insensitive to $\sstar$, varying from $\sdm=-0.23\pm0.60\,\csg$ for an assumed halo of $M_{200}=10^{13}M_\odot$ to $\sdm=-0.1\pm0.28\,\csg$ for a halo of $M_{200}=10^{15}M_\odot$.

\begin{figure}
\centering
\includegraphics[width=0.7\textwidth]{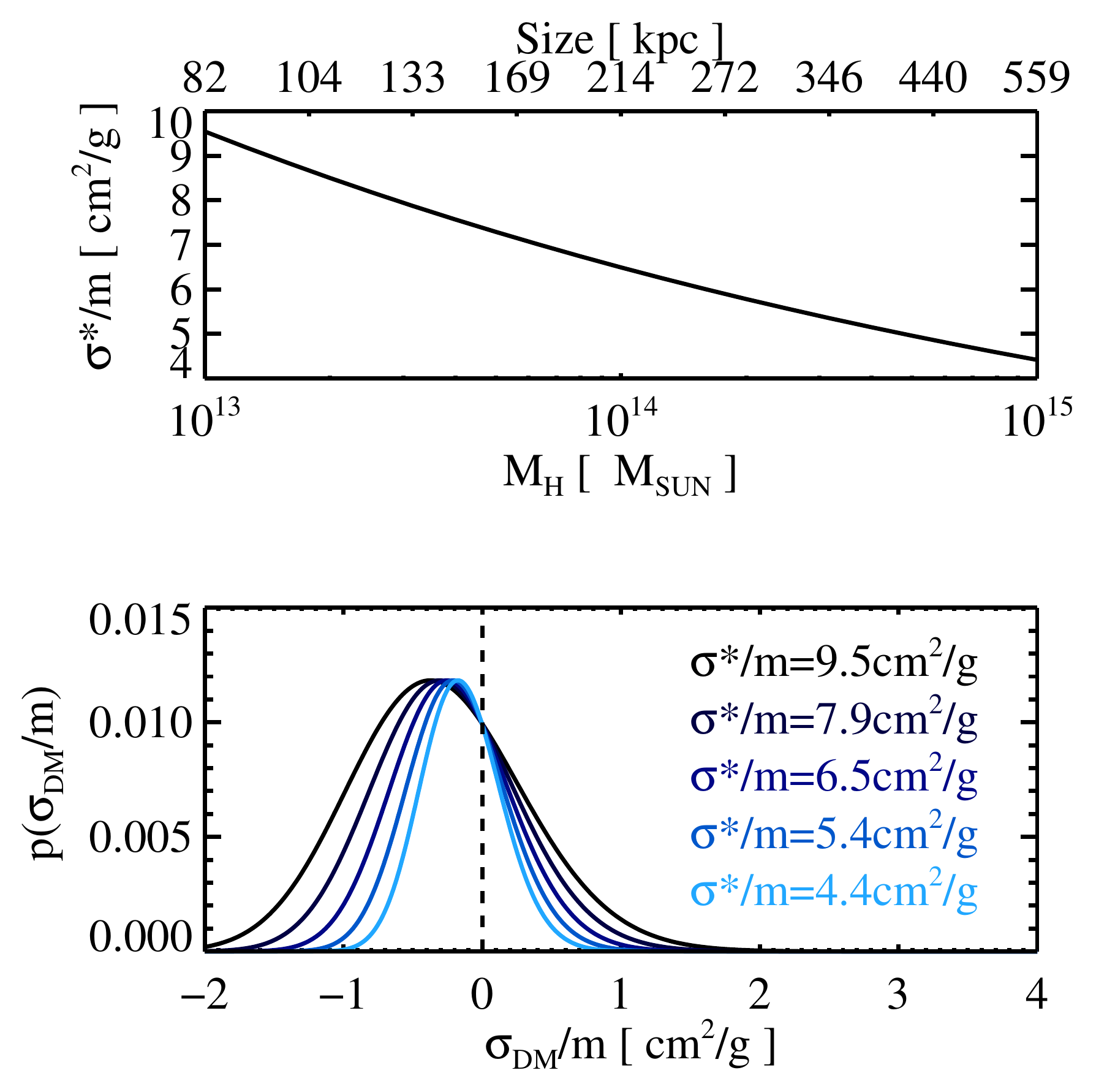}
\caption{The sensitivity of measurements of dark matter's self-interaction cross-section to the model parameter $\sstar$. This parameter is the characteristic value of cross-section at which an appropriately-sized cloud of standard model particles becomes optically thick. The top panel shows the value of $\sstar$ for different various substructure masses, assuming an NFW mass profile and a mass-concentration relation from cosmological simulations \cite{maccio}. The bottom panel demonstrates how a few of those values affect our measurement of the cross-section. The resulting variation is sub-dominant to statistical error in our sample of clusters. We adopted a value of $\sstar=6.5\pm3\,\csg$, corresponding to dark matter halos of $M=10^{14\pm1}M_\odot$, and propagated the uncertainty through to our final constraints.}
\label{fig:sigmastar}
\end{figure}

The relative behaviour of gas and dark matter was compared through a ratio in the prefactor
\begin{equation}
B=\frac{C_{\rm DM} A_{\rm DM} M_{\rm gas} \rho_{\rm DM}}{C_{\rm gas} A_{\rm gas} M_{\rm DM} \rho_{\rm gas}},
\end{equation}
where $C$, $A$ and $M$ are the drag coefficient, size and mass of the merging halo, and $\rho$ is the density of material through which it is moving. We assumed a conservative lower limit of $B\simgt1$, leading to a conservative upper limit on our constraints on $\sdm$. 

The first requirement to have ensured a conservative treatment is that the infalling substructure's gas envelope is smaller than its dark matter envelope, $A_{\rm gas}<A_{\rm DM}$. This is generically true of isolated structures in numerical simulations and, as gas is stripped during the collision, it will become smaller still. The geometric size of a gas halo also depends upon its temperature -- and hot gas may be more easily stripped than cold gas. To test whether that has an statistically significant effect, we measured the X-ray temperature of each observed infalling system, and separately analyzed the hotter and cooler halves of our sample. As shown in Figure~S7, the results for each half remain consistent, with error bars larger by approximately $\sqrt{2}$. For the hotter sample ($T>8$\,keV), we found $\sdm=-0.10\pm0.58\,\csg$ and for the cooler sample ($T<8$\,keV) we found $\sdm=-0.50\pm0.64\,\csg$. Although there was marginal evidence that hot gas is more easily stripped than cold gas, which could be investigated with a much larger sample, our conclusions remain unaffected within current statistical precision.

\begin{figure}
\centering
\includegraphics[width=0.7\textwidth]{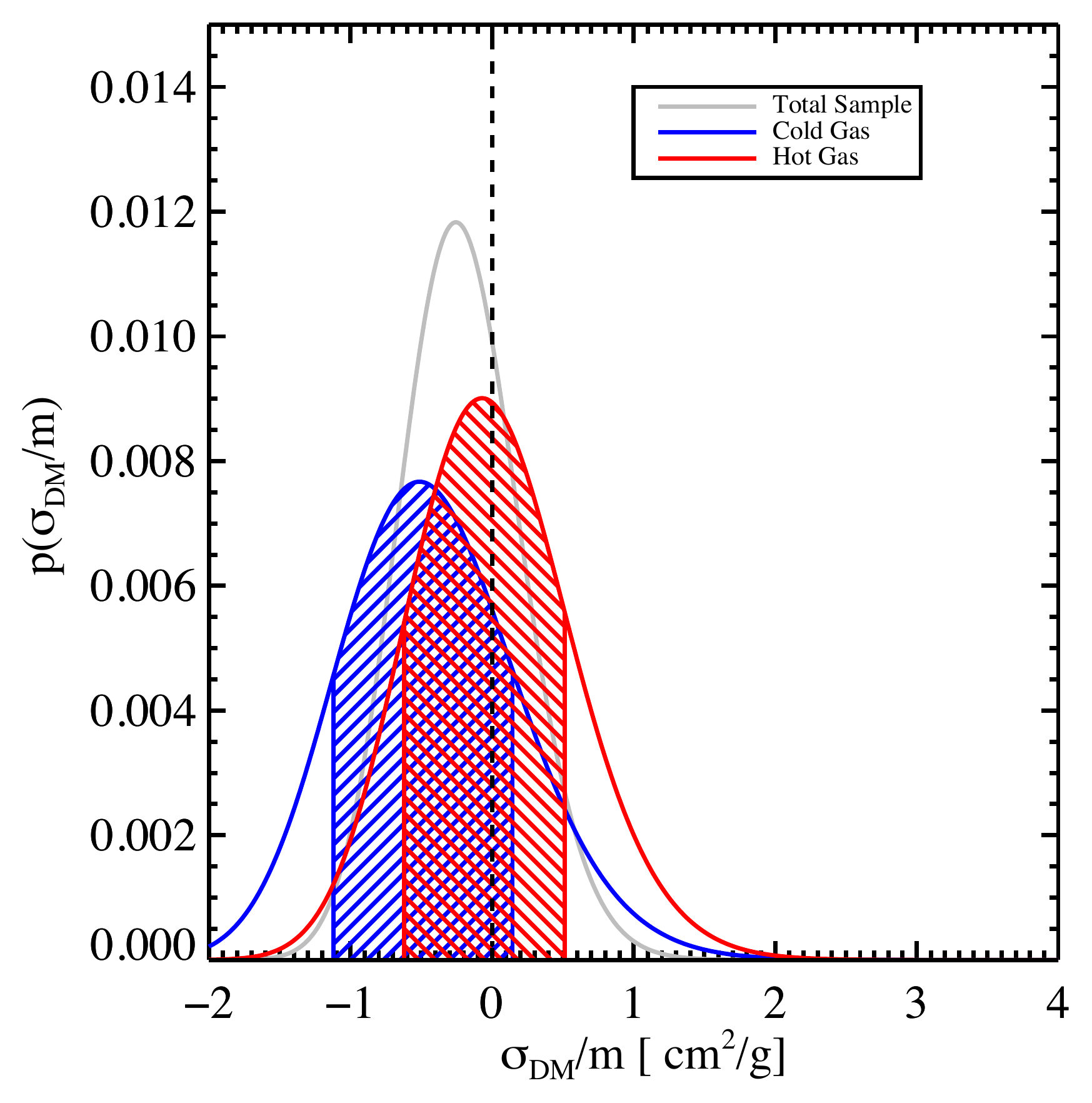}
\caption{The sensitivity of measurements of dark matter's self-interaction cross-section to the temperature of the gas against which dark matter's trajectory is calibrated. We measured the gas temperature from the X-ray spectra of our 72 systems, and split the sample in two: blue data show substructures with gas temperature $<8$\,keV, and red data show substructures with gas temperature $>8$\,keV. The constraining power of each sample is approximately $\sqrt{2}$ less than that of the full sample, shown in grey, and no statistically significant difference is measured.}
\label{fig:sensitivity}
\end{figure}

The second requirement to have ensured a conservative treatment is that the gas fraction in the medium through which the bullet is traveling, $f_{\rm gas}\equiv\rho_{\rm gas}/\rho_{\rm DM}$ is less than that of the infalling structure $M_{\rm gas}/M_{\rm DM}$. We assumed that, overall, infalling structure contains the universal fraction $\Omega_{\rm B}/\Omega_{\rm D}=0.184$ \cite{planckpars}, and we measured $f_{\rm gas}$ in mock data realised from cosmological simulations of structure formation \cite{baryonfracsimC}. The mean $f_{\rm gas}$ over all simulations (the solid line in Figure~S8) is lower than the universal fraction, and is indeed constant (within 10\%) at the radii of observable substructures (points in Figure~S8). These conclusions from simulations are consistent with deep X-ray observations of galaxy clusters, e.g.\ \cite{xraycosmopars}.

\begin{figure}
\centering
\includegraphics[width=0.7\textwidth]{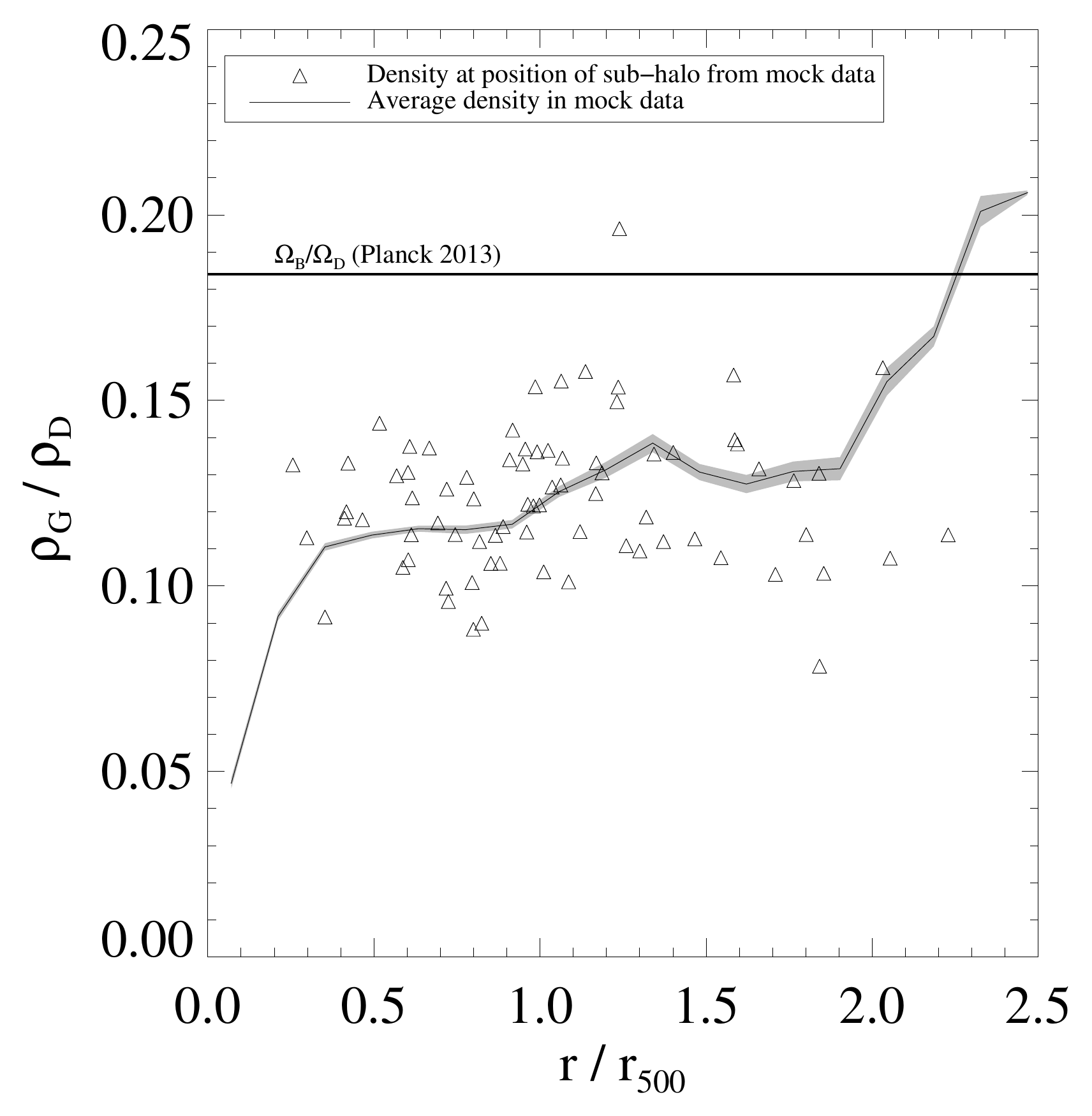}
\caption{The sensitivity of measurements of dark matter's self-interaction cross-section to the density of gas through which it is moving. The plot shows the gas fraction $f_{\rm gas} = \rho_{\rm gas} / \rho_{\rm DM}$ in simulated galaxy clusters \cite{baryonfracsimC}, as a function of clustercentric radius. The solid line shows the average  $f_{\rm gas}$ over 16 clusters, with the $1\sigma$ error on the mean given in grey. Triangles show the measured $f_{\rm gas}$ at the radius of substructures observable in mock 2D realisations of the 3D simulations (only the inner $\sim60\%$ lie inside the HST field of view at the redshifts of the observed systems). Our interpretation of the dark matter and gas trajectories as an interaction cross-section, assumes that these are lower than the universal fraction $\Omega_{\rm B}/\Omega_{\rm D}=0.184$ \cite{planckpars}.}
\label{fig:fgas}
\end{figure}

%DH analysed the optical imaging. DH and ET analysed the X-ray imaging. DH and RM measured the gravitational lensing. TK and AT developed the statistical tests to interpret the data. All authors contributed to the manuscript.

\clearpage
\bibliography{bibliography}
\bibliographystyle{Science}

\end{document}